\documentclass[aps,twocolumn,groupedaddress,amsmath,amssymb,nofootinbib]{revtex4-2}
\usepackage{graphicx}
\usepackage{color}

\begin{document}

\title{Non-linear charged planar black holes in four-dimensional Scalar-Gauss-Bonnet theories}

\author{Mois\'es Bravo-Gaete}
\email{mbravo-at-ucm.cl} \affiliation{Facultad de Ciencias
B\'asicas, Universidad Cat\'olica del Maule, Casilla 617, Talca,
Chile.}

\author{Luis Guajardo}
\email{lguajardo-at-ucm.cl} \affiliation{Facultad de Ciencias
B\'asicas, Universidad Cat\'olica del Maule, Casilla 617, Talca,
Chile.}

\author{Julio Oliva}
\email{julioolivazapata-at-gmail.com} \affiliation{Departamento de F\'isica, Universidad de Concepci\'on,
Casilla, 160-C, Concepci\'on, Chile.}

\begin{abstract}
In this work, we consider the recently proposed well-defined theory that permits a healthy $D\to 4$ limit of the Einstein-Gauss-Bonnet combination, which requires the addition of a scalar degree of freedom. We continue the construction of exact, hairy black hole solutions in this theory in the presence of matter sources, by considering a nonlinear electrodynamics source, constructed through the Pleba\'nski tensor and a precise structural function $\mathcal{H}(P)$. Computing the thermodynamic quantities with the Wald formalism, we identify a region in parameter space where the hairy black holes posses well-defined, non-vanishing, finite thermodynamic quantities, in spite of the relaxed asymptotic approach to planar AdS. We test its local stability under thermal and electrical fluctuations and we also show that a Smarr relation is satisfied for these black hole configurations.
\end{abstract}

\maketitle
\newpage

\section{Introduction}

Undoubtedly, General Relativity (GR) is the most successful tested theories of gravity, showing compatibility from experimental observations on the solar system to external astrophysical systems \cite{Will:2014bqa,Will:2014kxa}. In spite of such success, there are strong reasons to explore theories beyond GR. In fact, the tension between GR and Quantum Mechanics as well as the accelerated expansion of the universe \cite{SupernovaSearchTeam:1998fmf,SupernovaCosmologyProject:1998vns}, have triggered a high interest in the community to explore alternative gravity theories.

Beyond dimension four, Lovelock gravity defines an interesting conservative extension of GR \cite{Lanczos:1938sf,Lovelock:1971yv,Stelle:1976gc,Stelle:1977ry}. Indeed, it is the most general model constructed on the Riemann curvature tensor with second-order equations of motion. In three dimensions, it coincides with GR with a cosmological constant. In the four-dimensional case, Lovelock gravity adds a quadratic term, called the Gauss-Bonnet density,
\begin{equation}\label{LGB}
\mathcal{L}_{GB}=R^{2}-4 R_{\mu \nu }R^{\mu \nu}+ R_{\mu \nu \sigma \rho}R^{\mu \nu \sigma \rho},
\end{equation} 
but according to the Chern theorem \cite{chern}, the contribution of $\mathcal{L}_{GB}$ is proportional to the bulk term of the Euler characteristic of the spacetime manifold. Consequently, Lovelock gravity actively modifies GR in dimensions higher than four, only. In consequence, the simplest nontrivial extension, for $D\geq 5$, is captured by the action
\begin{equation}\label{EHGB}
S_{\tiny{EGB}}=\frac{1}{2}\int d^{D}x \sqrt{-g}\left[R-2\Lambda +\alpha \mathcal{L}_{GB}\right],
\end{equation} commonly referred to as the Einstein-Gauss-Bonnet model (EGB), which possesses a rich family of solutions, e.g., exact spherically symmetric black holes \cite{Boulware:1985wk,Wiltshire:1985us,Wheeler:1985nh,Wheeler:1985qd}, topological black holes \cite{Cai:1998vy,Cai:2001dz} as well as charged and/or hairy configurations via the addition of matter sources as Maxwell fields or non minimally coupled scalar fields \cite{Cvetic:2001bk,Maeda:2008ha,BravoGaete:2019rci,BravoGaete:2013acu}. Furthermore, the EGB model is also interesting for theoretical reasons. Recently, microscopic wormholes have been studied in \cite{Giribet:2019dmg, Chernicoff:2020tvr}, where $\alpha$ is related to the throat of the wormhole (see also \cite{Garraffo:2007fi}-\cite{Camanho:2013uda}). From a holographic perspective, particularly in applications on the transport coefficients,  
on planar manifolds the model induces a violation in the well known Kovtun-Son-Starinets (KSS)-bound, a universal limit that was proposed on the ratio between the shear viscosity $\eta$ and the entropy density $s$ ($\eta/s$)  \cite{Brigante:2007nu}. Also, the interpretation of higher-order curvature terms as $\alpha'$ corrections in the low energy limit of string theory (see for example \cite{Gross:1986iv,Metsaev:1987zx,Zwiebach:1985uq,Candelas:1985en,Green:1997di,Camanho:2014apa}) reinforces the interest on such terms, which as mentioned contribute to the dynamics only above dimension four.  

Recently, novel attempts to obtain a non-trivial quadratic contribution in the curvature, leading to second order field equations, even in dimension four have been explored. One of the consistent approaches comes from the ideas in ref. \cite{Hennigar:2020lsl, Fernandes:2020nbq, Hennigar:2020fkv}, where the regularization of the $D\rightarrow 4$ limit is due to a counterterm introduced through a conformal transformation, which generates a perfectly healthy four dimensional limit at the cost of an additional scalar degree of freedom. A second idea was constructed from a Kaluza-Klein reduction \cite{Lu:2020iav,Kobayashi:2020wqy} of a maximally symmetric internal spacetime of $D-4$ dimensions. On planar manifolds these two approaches converge to the same action principle\footnote{Interestingly enough, the exploration of these ideas was triggered by an ill-defined limit that is consistent only on four-dimensional spacetimes with symmetries as constructed in \cite{Glavan:2019inb} (see \cite{Gurses:2020ofy,Gurses:2020rxb} and \cite{Fernandes:2022zrq} for a detailed analysis of the inconsistency of the original approach and a review on this topic, respectively, and \cite{Aoki:2020lig} for an alternative approach of taking the $D\to 4$ limit by breaking the temporal diffeomorphism invariance).}.

The relevant action principle in four dimensions reads:
\begin{eqnarray}
S&=&\int d^{4}x \sqrt{-g} \mathcal{L}_{g}\nonumber\\
&=&\frac{1}{2}\int d^{4}x \sqrt{-g} \Big[ R-2 \Lambda +{\tilde{\alpha}}(\phi \mathcal{L}_{GB}+ 4 G^{\mu \nu} \phi_{\mu} \phi_{\nu}\nonumber\\
&-&4 X \Box \phi+ 2 X^{2})\Big].\label{eq:EH-GB-reg}
\end{eqnarray}
Here, $G^{\mu \nu}$ is the Einstein tensor, and for simplicity we adopt the following notation:
$\phi_{\mu}:=\nabla_{\mu} \phi$, $\Box \phi= \nabla^{\sigma} \nabla_{\sigma} \phi$, and $X:=\nabla_{\mu} \phi \nabla^{\mu} \phi$ stands for the kinetic term. It is noted that the theory under consideration belongs to the shift-symmetric sector of the Horndeski family \cite{Horndeski:1974wa}, and in terms of the Galileon theory it is constructed via a covariant formulation (see for ex. Refs. \cite{Nicolis:2008in}-\cite{Deffayet:2011gz}). Hereafter we will refer to the action (\ref{eq:EH-GB-reg}) as \emph{Four dimensional-scalar-Einstein-Gauss-Bonnet} (4DS-EGB), and it is the theory that we will consider throughout this work.

This theory has been intensively studied in the recent years: black holes with spherical topology were constructed in \cite{Lu:2020iav,Glavan:2019inb}, together with the inclusion of axionic fields for planar black holes in \cite{Wang:2020uiq}. Charged solutions in the context of the non-linear electrodynamics of Born-Infeld were constructed in  \cite{Meng:2021huz}, and the properties of compact objects were studied in \cite{Charmousis:2021npl}. Even three dimensional scenarios for static \cite{Hennigar:2020fkv} and spinning configurations \cite{Hennigar:2020drx,Ma:2020ufk,Konoplya:2020ibi} have been explored, obtaining a generalization of the Ba\~{n}ados-Teitelboim-Zanelli black hole \cite{Banados:1992wn}, as well as the Bondi-Sachs framework \cite{Lu:2020mjp}.

In the present work, continuing the investigations of 4DS-EGB coupled to matter, we introduce a further scalar $\psi$, which is conformally coupled, as well as a non-linear electrodynamics, namely we consider the action
\begin{equation}\label{eq:action}
S[g_{\mu\nu},\phi,A_{\mu}, P^{\mu\nu}]=S+S_{\psi}+S_{NLE},
\end{equation}
where
\begin{eqnarray*}
S_{\psi} &=&\int{d}^4x\sqrt{-g} {\cal{L}}_{\psi}\\
&=& \int{d}^4x\sqrt{-g}\Biggl[
-\frac{1}{2}\partial_{\mu}\psi\partial^{\mu}\psi-\frac{1}{12}R\psi^2-\zeta \psi^4\Biggr],\nonumber\\
\end{eqnarray*}
and
\begin{eqnarray*}
S_{NLE} &=& \int{d}^4x\sqrt{-g} {\cal{L}}_{NLE}\\
&=&\int{d}^4x\sqrt{-g}\Biggl[-
\frac{1}{2}F_{\mu\nu}P^{\mu\nu} + \mathcal{H}(P)\Biggr].\nonumber\\
\end{eqnarray*}

\textcolor{black}{In general, the electrodynamics is described by the field strength $F_{\mu\nu}=\partial_{\mu}A_{\nu} - \partial_{\nu}A_{\mu}$ and its Hodge dual, $\star F_{\mu\nu}$, tensors that allow to construct a scalar $F=\dfrac{1}{4}F_{\mu\nu}F^{\mu\nu}$ and a pseudo-scalar $G=\dfrac{1}{4}\star F_{\mu\nu}F^{\mu\nu}$. If $L=L(F,G)$ is the lagrangian density of the non-linear electrodynamic source, the Plebánski tensor $P_{\mu\nu}$ is formally defined through
\begin{equation}
P_{\mu\nu} \equiv 2\dfrac{\partial L}{\partial F_{\mu\nu}} = L_F F_{\mu\nu} + L_G \star F_{\mu\nu},
\end{equation}
where $L_F,L_G$ are partial derivatives. Using a Legendre transformation, the lagrangian can be successfully rewritten using a structural function $\mathcal{H}$ (a Hamiltonian function), which in general depends on $P=\dfrac{1}{4}P_{\mu\nu}P^{\mu\nu}$ and $Q=\dfrac{1}{4}\star P_{\mu\nu}P^{\mu\nu}$, but in this work we will focus on static configurations, so that $\mathcal{H}=\mathcal{H}(P)$\cite{Plebanski:1970zz}. Here, we will deal with electrically charged configurations, $A_{\mu}= A_{t}(r)dt$, and for simplicity we will fix the structural function $\mathcal{H}(P)$ as:}

\begin{equation}\label{eq:H}
\mathcal{H}(P)= a_1 \sqrt{-2{P}} + a_2{P},
\end{equation}
where $\zeta, a_1,a_2$ are coupling constants that will be suitably fixed. As shown below, this model allows for planar, asymptotically AdS, exact black hole solutions, which approach the background in a relaxed manner as compared with the asymptotic conditions constructed by Henneaux-Teitelboim in \cite{Henneaux:1985tv}. In spite of such slow asymptotic behavior, we show that using the Wald's method to compute Noether charges leads to a finite expression for the mass, the entropy and the electric charge, which even more fulfill an Smarr-type relation.  

It it known that non-minimally coupled, self-interacting scalars with self interactions allow to construct hairy black holes as well a other interesting configurations in vacuum in GR (see e.g. \cite{Anabalon:2012tu}-\cite{Barrientos:2022yoz}). Even more, quartic self interactions allow to construct hairy black holes in the presence of higher-curvature terms in dimension greater than four,  $\psi$ \cite{BravoGaete:2013djh, Correa:2013bza}, which has led us to considering the 4DS-EGB model in presence of such a further scalar degree of freedom. Most of the solutions considered in the latter references lead to vanishing charges, which as explained in \cite{Bravo-Gaete:2021hza} can be  circumvented e.g. by the introduction of a power-law Maxwell term. Even more, it is known that conformally coupled scalar fields in dimension four allow to by-pass no-hair results as shown in the pioneering works leading to the BBMB solution \cite{Bocharova:1970skc, Bekenstein:1974sf} (see also \cite{Martinez:2002ru, Martinez:2004nb} for its extension in the presence of a non-vanishing cosmological constant as well as  \cite{Bardoux:2012tr} and \cite{Cisterna:2018hzf} for extra dressings of such solutions given by a Kalb-Ramond potentials and a massless scalar fields which is linear along the planar coordinates of the horizon, respectively). On the other hand on the electromagnetic side, recently, there has been a growing interest in a different nonlinear electrodynamics source, characterized by the antisymmetric conjugate tensor $P_{\mu \nu}$ (the Pleba\'nski tensor), and a structure-function $\mathcal{H}(P)$. This choice of a matter source has allowed obtaining regular black holes solutions \cite{Ayon-Beato:1998hmi,Ayon-Beato:1999qin,Ayon-Beato:1999kuh,Ayon-Beato:2000mjt,Ayon-Beato:2004ywd}, an exact solution of a massive, electromagnetically charged and rotating configuration \cite{Garcia-Diaz:2021bao,Diaz:2022roz,Ayon-Beato:2022dwg}, black holes with non-standard asymptotic behavior \cite{Alvarez:2014pra,Zhu:2020zti}, slowly rotating black holes \cite{Kubiznak:2022vft}, and the construction of black holes within the context of Critical Gravity \cite{Alvarez:2022upr}. A hint to the present work is taken from this last reference. As it is known, Critical Gravity generically leads to fourth-order equations of motion, and its vacuum solution from ref. \cite{Lu:2011zk} admits an AdS black hole with vanishing thermodynamic quantities. A complete analysis of this statement is performed through Noether-Wald charges in \cite{Anastasiou:2017rjf, Anastasiou:2021tlv}, also for its six-dimensional analog. The introduction of the NLE source in \cite{Alvarez:2022upr} generates a fruitful interaction with all the integration constants, leading to a stable black hole with non-null mass, entropy, and electric charge.

As a consequence, there is enough evidence to conjecture that the 4DS-EGB theory, coupled to a conformal scalar and interacting with a non-minimal electrodynamics, may lead to exact hairy, charged black hole solutions. The presence of the NLE leads to non-vanishing, finite charges. In the present work we confirm such expectation.

This paper is structured as follows: In Section \ref{Section-derivation} we will derive the four-dimensional solution to be discussed, taking into account an arbitrary value of the constant $\tilde{\alpha}$. In Section \ref{Section-termo}, we explore the thermodynamics of these configurations via the Wald formalism, studying their local stability under thermal fluctuations and electrical fluctuations respectively. Finally, Section \ref{Section-conclusions} is devoted to our conclusions and further discussion.

\section{The setup and the solution}\label{Section-derivation}
The field equations of the model considered here, with action principle given in \eqref{eq:action} are:
\begin{eqnarray}
&&{\cal{E}}_{\mu \nu}:=G_{\mu \nu}+\Lambda g_{\mu \nu}-\tilde{\alpha} T_{\mu \nu}^{\phi}-T_{\mu \nu}^{\psi} - T_{\mu\nu}^{NLE}=0 \label{eq:Emunu},\\
&&{\cal{E}}_{\phi}:={\mathcal{L}}_{GB}-\nabla_{\mu}\left(8 G^{\mu \nu}\phi_{\nu}-8\Box\phi \phi^{\mu}+8 X \phi^{\mu}\right) \nonumber \\
&&\hspace*{1cm}-4\nabla_{\nu}\nabla_{\mu}\left(X g^{\mu \nu}\right)=0,\label{eq:Ephi}\\
&&{\cal{E}}_{\psi}:=\Box \psi-\dfrac{1}{6} R\psi- 4\zeta \psi^3=0,\label{eq:Epsi} \\
&&{\cal{E}}^{\nu}_{F}:= \nabla_{\mu}P^{\mu\nu} = 0,\label{eq:EF} \\
&&{\cal{E}}^{\mu \nu}_{P}:= \left( \dfrac{\partial {\cal{H}}}{\partial P}\right) P^{\mu\nu}-F^{\mu\nu}  = 0,\label{eq:EP}
\end{eqnarray}
where $G_{\mu\nu}$ is the Einstein tensor, and the explicit form of the energy-momentum tensors $T_{\mu \nu}^{\phi}, T_{\mu \nu}^{\psi}, T_{\mu\nu}^{NLE}$  are given in the appendix.

To begin the derivation of the solution, we consider the following ansatz
\begin{eqnarray}\label{eq:metric}
ds^2 = -f(r)~dt^2 + \dfrac{dr^2}{f(r)} + r^2d\Omega_{2}^{2},
\end{eqnarray}
with $d\Omega_{2}^2$ the line element for the Euclidean flat space of dimension 2, where we assume that the planar coordinates belong to a compact set $0 \leq x_1 \leq \Omega_{x_1}$ and $0 \leq x_2 \leq \Omega_{x_2}$. The difference $\mathcal{E}_{t}^{t}-\mathcal{E}_{r}^{r}=0$ provides the branch $\phi(r)=\ln(r)$ followed up by 
\begin{equation}\label{eq:psi}
\psi(r)=\dfrac{\sqrt{a}}{r},
\end{equation}
where $a$ is a positive integration constant. It is noted that the logarithmic behavior of $\phi(r)$ ensures that ${\cal{E}}_{\phi} =0 $ is satisfied.

The equation ${\cal{E}}_{\psi}=0$ leads to a non-homogeneous second-order Euler differential equation for the metric function $f(r)$ given by
\begin{eqnarray}
&&r^2 f''-2rf'+2f=24\zeta a, \nonumber
\end{eqnarray}
where $(')$ denotes the derivative with respect to the radial coordinate $r$, and the general expression takes the form
\begin{eqnarray}\label{eq:f}
f(r)&=&C_2r^2 + C_1r + 12\zeta a,
\end{eqnarray} 
where $C_1$ and $C_2$ are integration constants. The Maxwell equation ${\mathcal{E}}_F = 0$ leads to 
\begin{eqnarray}\label{eq:Prt}
P^{rt}= \frac{Q}{r^2},
\end{eqnarray}
so that ${P}= -{Q^2}/{(2r^4)}$ and, by using $\mathcal{E}_P=0$, one obtains:
\begin{equation}\label{eq:Frt}
 F^{rt} = a_2\left(\dfrac{Q}{r^2}\right) -a_1.
\end{equation}

With the above setup, in order to solve the rest of the Einstein equations some relations between the constants will emerge. In fact, the analysis of the equations leads to a black hole solution described by a single integration constant. With this in mind, we will opt for the following representation of the metric function:
\begin{equation}\label{eq:f}
f(r)= \dfrac{r^2}{2 \tilde{\alpha}} \left( 1 + \dfrac{(\mu-1)r_h}{r} - \dfrac{\mu r_h^2}{r^2} \right),
\end{equation}
to express all quantities in terms of $r_h$, which will be our degree of freedom that also represents the location of the horizon, since $f(r_h)=0$. The parameter $\mu$ in the metric function switches on/off the scalar field, but it is also related to both coupling constants of the structural function ${\cal{H}}({P})$. In fact, one gets the following set of relations:
\begin{eqnarray}
\label{sol:integrationconstants}a &=& -\dfrac{\mu r_h^2}{24 \tilde{\alpha} \zeta},\quad Q= -r_h^2,\quad \Lambda = -\dfrac{3}{4 \tilde{\alpha}}, \\
\label{sol:coupling_a}a_1&=& -\dfrac{(\mu-1)^2}{4 \tilde{\alpha}},\quad a_2 = -\dfrac{\mu^2 (144\zeta \tilde{\alpha} +1)}{288 \zeta \tilde{\alpha}^2}.
\end{eqnarray}

Many comments can be raised with respect to the solution (\ref{eq:H}), (\ref{eq:metric})-(\ref{eq:psi}), (\ref{eq:Prt})-(\ref{sol:coupling_a}). The first one, in order to obtain a real scalar field $\psi$ in (\ref{eq:psi}), from (\ref{sol:integrationconstants}) we have that the quotient between the constants $\mu$ and $\zeta$ must be negative (here we have supposed $\tilde{\alpha}>0$), together with a negative cosmological constant $\Lambda$. Additionally, the particular cases $\mu=1$ and $\mu=0$ allow us to explore two branches, where for the first situation the structural function becomes linear (${\cal H}(P)\sim P$) and the metric functions takes the form $f(r)=\frac{r^2}{2 \tilde{\alpha}} \left(1-\frac{r_h^2}{r^2}\right)$, while that for the uncharged situation (this is $\mu=1$ and $144 \zeta \tilde{\alpha}=-1$) the integration constant $a$ and the location of the event horizon $r_h$ are related as $a=6 r_h^2$, which can be considered as the four dimensional limit of the spacetime considered in \cite{BravoGaete:2013djh,Correa:2013bza}. On the other hand, it is interesting to note that when the constant $\mu$ vanishes, the scalar field $\psi$ becomes zero, while ${\cal H}(P)\sim \sqrt{-2P}$ together with $f(r)=\frac{r^2}{2 \tilde{\alpha}} \left(1-\frac{r_h}{r}\right).$

Notice that from \eqref{eq:f}, on can see that the spacetime approaches a locally AdS spacetime in the planar foliation. The subleading terms in the metric component are relaxed with respect to those of Henneaux-Teitelboim \cite{Henneaux:1985tv} since in our case
\begin{eqnarray}
g_{tt}&=-\frac{r^2}{l^2}+f_{tt}r+\mathcal{O}\left(1\right)\ ,\\
g^{rr}&=\frac{r^2}{l^2}+f_{rr}r+\mathcal{O}\left(1\right)\ ,
\end{eqnarray}
where the coefficients $f$ denote the terms that define an slow asymptotic approach.

Given the structure of this new four-dimensional charged hairy configuration, and the relation between the integration constants $r_h$, $a$ and $Q$, in the following section we will derive its thermodynamic quantities.

\section{Thermodynamics Analyisis via Wald Formalism}\label{Section-termo}

After obtaining the solution for the previous model, in the following lines we compute and analyze their thermodynamic quantities via the Wald formalism \cite{Wald:1993nt,Iyer:1994ys}, constructing a conserved Noether current. As a first step, the variation of the total action, namely eq. (\ref{eq:action}), is formally written as
\begin{eqnarray}
\delta S&=& \sqrt{-g}\big[{\cal{E}}_{\mu \nu} \delta g^{\mu \nu} + {\cal{E}}_{\phi} \delta \phi+ {\cal{E}}_{\psi} \delta \psi \nonumber \\
&+&{\cal{E}}^{\nu}_{F} \delta (A_{\nu})+{\cal{E}}^{\mu \nu}_{P} \delta (P_{\mu \nu}) \big]+\partial_{\mu} {\cal{J}}^{\mu}, \label{deltaS}
\end{eqnarray} 
where ${\cal {E}}_{\mu \nu} $ are the equations of motion with respect to the metric, while that ${\cal{E}}_{\phi}$, ${\cal{E}}_{\psi}$, ${\cal{E}}^{\nu}_{F}$ and ${\cal{E}}^{\mu \nu}_{P}$ are the field equations with respect to $\phi$, $\psi$, $A_{\nu}$  and $P_{\mu \nu}$ respectively, present in the equations (\ref{eq:Emunu})-(\ref{eq:EP}). Together with the above expressions, from the equation (\ref{deltaS}) a surface term $ {\cal{J}}^{\mu}$ arises, which reads
\begin{eqnarray}
{\cal{J}}^{\mu}&=&\sqrt{-g}\Big[2\left(P^{\mu (\alpha\beta)
\gamma}\nabla_{\gamma}\delta g_{\alpha\beta}-\delta g_{\alpha\beta} \nabla_{\gamma}P^{\mu(\alpha\beta)\gamma}\right) \nonumber \\
&+&\frac{\delta \cal{L}}{\delta (\phi_{\mu})} \delta
\phi-\nabla_{\nu}\left(\frac{\delta \cal{L}}{\delta (\phi_{\mu
\nu})}\right) \delta \phi
+\frac{\delta \cal{L}}{\delta (\phi_{\mu \nu})} \delta (\phi_{\nu}) \nonumber\\
&-&\frac{1}{2}\frac{\delta \cal{L}}{\delta (\phi_{\mu \rho})}
\phi^{\sigma} \,
\delta g_{\sigma \rho}-\frac{1}{2}\frac{\delta \cal{L}}{\delta (\phi_{\rho \mu})}\phi^{\sigma} \,\delta g_{\sigma \rho} \label{eq:surface}  \\
&+&\frac{1}{2}\frac{\delta \cal{L}}{\delta ( \phi_{\sigma
\rho})}\phi^{\mu}\,\delta g_{\sigma \rho}+\frac{\delta \cal{L}}{\delta (\psi_{\mu})} \delta
\psi+
\frac{\delta \cal{L}}{\delta ( \partial_{\mu} A_{\nu})} \delta
A_{\nu}
\Big] \nonumber ,
\end{eqnarray}
where ${\cal{L}}={\cal{L}}_{g}+{\cal{L}}_{\psi}+{\cal{L}}_{NLE}$ is the lagrangian for the total action, $\psi_{\mu}:=\nabla_{\mu} \psi$, and $P^{\alpha \beta \gamma \delta}$, ${\delta \cal{L}}/{\delta (\phi_{\mu})}$,
${\delta \cal{L}}/{\delta (\psi_{\mu})}$, ${\delta \cal{L}}/{\delta (\phi_{\mu \nu})}$ and
${\delta \cal{L}}/{\delta (\partial_{\mu} A_{\nu})}$ are reported in the Appendix. To compute the thermodynamic quantities using the surface term given in (\ref{eq:surface}), we first define a $1$-form ${\cal{J}}_{(1)}={\cal{J}}_{\mu} dx^{\mu}$ as well as its Hodge dual ${\Theta}_{(3)}=(-1)*{\cal{J}}_{(1)}$. Then, after making use of the equations of motions, we have the expression 
$${\cal{J}}_{(3)}={\Theta}_{(3)}-i_{\chi}*\mathcal{L}=-d* {\cal{J}}_{(2)},$$
where $i_{\chi}$ is a contraction of the vector field $\chi^{\mu}$ on the first index of $*\mathcal{L}$. The above relation allows to define a $2$-form ${Q}_{(2)}=*{\cal{J}}_{(2)}$ such that ${\cal{J}}_{(3)}=d Q_{(2)}$, which in this case takes the following form:
\begin{eqnarray}
Q_{(2)}&:=& Q_{\alpha_1 \alpha_2} \label{eq:noether} \\ 
&=&\varepsilon_{\alpha_1 \alpha_2 \mu \nu}\Big[2P^{\mu\nu\rho\sigma}\nabla_\rho \chi_\sigma -4\chi_\sigma
\nabla_\rho P^{\mu\nu\rho\sigma}\nonumber \\
&+&\frac{\delta \cal{L}}{ \delta \phi_{\mu \sigma}} \phi^{\nu} \chi_{\sigma}-\frac{\delta \cal{L}}{ \delta \phi_{\nu \sigma}} \phi^{\mu}
\chi_{\sigma}-\frac{\delta \cal{L}}{\delta ( \partial_{\mu} A_{\nu})} \chi^{\sigma} A_{\sigma}\Big]. \nonumber
\end{eqnarray}
We report in the Appendix the explicit expression for each element from (\ref{eq:noether}). The vector field $\chi^{\mu}$ is supposed to be a time-translation vector, which is a Killing vector and it is null on the location of the event horizon where, as before, is denoted as $r_h$.
Finally, the variation of the Hamiltonian reads
\begin{eqnarray}
\delta \mathcal{H}&=&\delta \int_{\mathcal{C}} {\cal{J}}_{(3)} -\int_{\mathcal{C}} d \left(i_{\chi} \Theta_{(3)}\right) \nonumber \\
&=& \int_{\Sigma^{(2)}}\left(\delta {Q}_{(2)}-i_{\chi} {\Theta}_{(3)}\right),\label{var_diff}
\end{eqnarray}
where $\mathcal{C}$ and $\Sigma^{(2)}$ represent a Cauchy Surface and its boundary respectively. Here we note that (\ref{var_diff}) has two components, one of them located at infinity (denoted as $ \mathcal{H}_{\infty}$) and the other at the horizon (given by $\mathcal{H}_{+}$).

In our case, the boundary term reads 
\begin{equation}
 \delta {\cal{H}} = - \dfrac{3}{2}\dfrac{\mu(\mu-1)\left(\zeta\tilde{\alpha} + \frac{1}{432}\right)}{\tilde{\alpha}^2 \zeta} r_h^2 \delta r_h \Omega_2,
\end{equation}
where $\Omega_2$ is the finite volume of the compact planar base manifold given by $\int dx_1 dx_2=\int d \Omega_{2}=\Omega_2=\Omega_{x_1} \Omega_{x_2}$, so that the contribution at the infinity is related to the mass parameter, which reads
\begin{equation}\label{eq:M}
{\cal{M}} = -\dfrac{\mu(\mu-1)}{2\tilde{\alpha}^2 \zeta} \left(\zeta\tilde{\alpha} + \frac{1}{432}\right)r_h^3 \Omega_{2}. 
\end{equation}
We immediately note that, in order to obtain a non-null mass for the black hole, we need as a minimum to turn on the scalar field $\psi$ and to receive both contributions from our structural function (\ref{eq:H}). This is, in turn, the same observation that led to nonzero thermodynamic quantities in four-dimensional Critical Gravity \cite{Alvarez:2022upr} as well as the Einstein-Proca model \cite{Alvarez:2014pra,Zhu:2020zti}, and that highlights the importance of the choice of the NLE as a matter source.

On the other hand, the component at the horizon reads $$\delta {\cal{H}}_{+} = T\delta{\cal{S}} + \Phi_e \delta{\cal{Q}}_e,$$
an in order to construct the rest of the thermodynamic quantities, we start with the Hawking temperature for this solution, which reads
\begin{equation}
T=\dfrac{(\mu+1)r_h}{8\tilde{\alpha} \pi}, \label{eq:T}
\end{equation}
and the electric potential is defined as 
\begin{equation}\label{eq:Phi}
\Phi_e = -A_t(r_h) = (a_1+a_2)r_h,
\end{equation}  where the constants $a_1$ and $a_2$ was given previously in (\ref{sol:coupling_a}), and the electric charge takes the form
\begin{equation}\label{eq:Q}
 {\cal{Q}}_e= -Q=r_h^2 \Omega_{2}. 
\end{equation}
Finally, the Wald entropy reads
\begin{equation}\label{eq:S}
{\cal{S}} = \dfrac{\pi(\mu+144\zeta \tilde{\alpha})}{72 \zeta \tilde{\alpha}} r_h^2 \Omega_2.
\end{equation}

\begin{figure}[!ht]
\begin{center}
\includegraphics[scale=0.8]{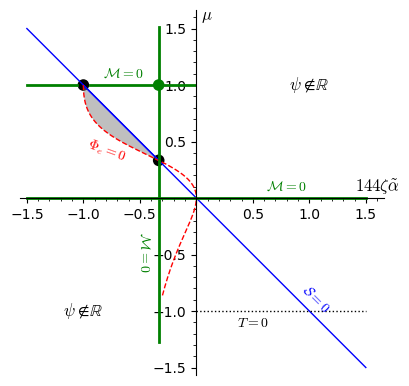}
\caption{{Graphic representation of the region ${\cal{R}}$ (filled, in gray) where the constants $\zeta$ and $\mu$ can be chosen in order to satisfy that all the thermodynamic quantities of the solution (\ref{eq:H}),(\ref{eq:metric})-(\ref{eq:psi}), (\ref{eq:Prt})-(\ref{sol:coupling_a}) are non-negative. We scale the axis to be $144\zeta \tilde{\alpha}$ and $\mu$ for simplicity. The big black dot located at $(-1,1)$ in our graph, corresponds to the four-dimensional limit from refs. \cite{BravoGaete:2013djh, Correa:2013bza}. A nontrivial interaction between the nonlinear source and the geometry is observed at $\left(-\frac{1}{3},\frac{1}{3}\right)$}.}
\label{fig:termo}
\end{center}
\end{figure}

With respect to these thermodynamic parameters, we can mention that for a suitable choice of the constants $\mu$ and $\zeta$, it is possible to obtain positive expressions for the extensive as well as the intensive parameters, as it is shown in Figure \ref{fig:termo}. The zero-entropy condition is plotted in the diagonal line (blue), and the entropy is positive in the bottom half region, $\mu+144\zeta\tilde{\alpha}<0$. The zero-mass conditions (green) are represented with horizontal lines at $\mu=0$ and $\mu=1$, and the vertical line at $\zeta = -\frac{1}{432\tilde{\alpha}}$ (or $144\zeta\tilde{\alpha}=-\frac{1}{3}$ in the scale chosen for the plot). The electric potential vanishes along the dashed curve (red). Recall that the scalar field $\psi$ is real provided $\mu\zeta <0$ (\ref{sol:coupling_a}), which forces us to work in the second and fourth quadrants. As it was shown before, for the special case $\mu=1$ the structural function ${\cal{H}}(P)$ is linear (\ref{eq:H}). In this case, the mass ${\cal{M}}$ vanishes, but ${\cal{S}} >0$ and $\Phi_e < 0$ when $\zeta < -\frac{1}{144 \tilde{\alpha}}$, which is represented to the left of the big black point at $(-1,1)$. Indeed, this special point corresponds to the four-dimensional limit of refs. \cite{BravoGaete:2013djh, Correa:2013bza}. Surprisingly, our analysis unveils another point where the mass and the entropy vanishes, obtained when $\mu=\frac{1}{3}, 144\zeta\tilde{\alpha}= -\frac{1}{3}$. Remarkably, the electric potential also vanishes in that point, even when the scalar field and the nonlinear source are non trivially interacting with the geometry.

Moving to the right of the point $(-1,1)$, the big point at $\left(-\frac{1}{3},1\right)$ (colored in green) represents the case when $a_2=1$, which recovers the classical Maxwell source, since $P_{\mu\nu}=F_{\mu\nu}$ in that specific case. It is noted that, in consistency with the results from ref. \cite{Bravo-Gaete:2021hza}, the entropy becomes negative in this sector. In contrast, when $\mu=0$ the scalar field $\psi$ vanishes and ${\cal H}(P)\sim \sqrt{-2P}$, where ${\cal{M}}=0$, ${\cal{S}}=2 \pi r_h^2 \Omega_{2}$, while that the Hawking temperature $T$ and the electric potential $\Phi_{e}$ are given by $T=r_h/( 8 \pi \tilde{\alpha})$ and $\Phi_e=-r_h/(4\tilde{\alpha})$ respectively. Finally, the dotted line at $\mu=-1$ corresponds to an extremal solution, in the sense that the Hawking temperature vanishes (\ref{eq:T}), contrary to the other quantities which are not null (see (\ref{eq:M}), (\ref{eq:Phi})-(\ref{eq:S})). In this extremal solution, the entropy ${\cal{S}}$ can be positive if we choose $\zeta \in \left]0,\frac{1}{144 \tilde{\alpha}}\right[$.
\begin{figure}[!ht]
\begin{center}
\includegraphics[scale=0.55]{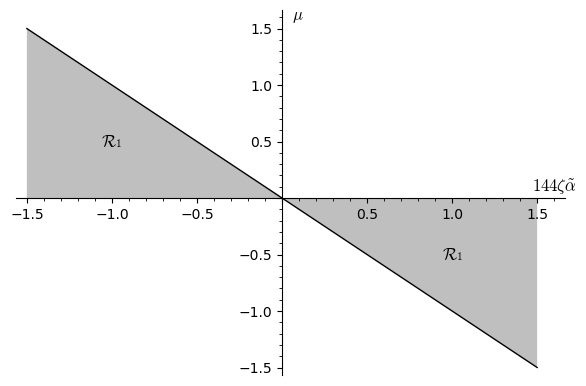}
\caption{{Graphic representation of the region ${\cal{R}}_{1}$, where the constants $\zeta$ and $\mu$ satisfy the condition $C_{\Phi_{e}} \geq 0$.}}
\label{fig:ce}
\end{center}
\begin{center}
\includegraphics[scale=0.55]{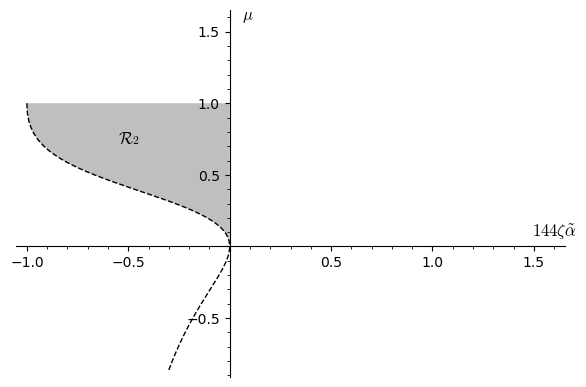}
\caption{{Graphic representation of the region ${\cal{R}}_{2}$, where the constants $\zeta$ and $\mu$ satisfy the condition $\epsilon_{T} > 0$.}}
\label{fig:et}
\end{center}
\end{figure}

We end this section by analyzing this charged black hole configuration as a thermodynamic system under small perturbations around the equilibrium, we will consider {\em{the grand canonical ensemble}}, where the temperature $T$ as well as the electric potential $\Phi_e$ are fixed quantities. As a first step, the mass ${\cal{M}}$, the entropy ${\cal{S}}$, and the electric charge ${{\cal{Q}}_{e}}$ can be rewritten in the function of these intensive thermodynamic parameters as follows 
\begin{eqnarray}
{\cal{M}}&=&\frac{16 \pi^3 T^3 \tilde{\alpha} (144 \zeta \tilde{\alpha}+\mu) \Omega_2}{27 (\mu+1)^2 \zeta}
+ \frac{2 \Phi_e^2 \Omega_2}{3 (a_1+a_2)^2} \label{eq:mass},\\
{\cal{S}}&=&\frac{8 \pi^3 T^2 \tilde{\alpha} (144 \zeta \tilde{\alpha}+\mu) \Omega_2}{9 (\mu+1)^2 \zeta},\\
{\cal{Q}}_{e}&=& \frac{\Phi_e^2 \Omega_2}{(a_1+a_2)^2} \label{eq:charge},
\end{eqnarray}
where the local thermodynamic (in)stability {under thermal fluctuations} can be determined via the behavior of the specific heat $C_{\Phi_{e}}$, which reads
\begin{eqnarray}\label{eq:cq}
C_{\Phi_{e}}&=&\left(\frac{\partial \mathcal{M}}{\partial T}\right)_{\Phi_{e}}=T \left(\frac{\partial \mathcal{S}_{W}}{\partial T}\right)_{\Phi_{e}}\nonumber\\
&=&\frac{16 \pi^3 T^2 \tilde{\alpha} (144 \zeta \tilde{\alpha}+\mu) \Omega_2}{9 (\mu+1)^2 \zeta},
\end{eqnarray}
where the sub-index stands for a constant electric charge ${\Phi_{e}}$. From (\ref{eq:cq}), in order to have a non-negative expression, we need to consider the special case 
$$\frac{\tilde{\alpha} (144 \zeta \tilde{\alpha}+\mu) }{\zeta (\mu+1)^2} \geq 0,$$
which is satisfied considering the region ${\cal{R}}_{1}$ from Figure \ref{fig:ce}, allowing us a locally stable configuration under thermal fluctuations. Together with the above, the study of charged configurations allows the analysis of how its response now under electrical fluctuations, characterized through the electric permittivity $\epsilon_{T}$  \cite{Gonzalez:2009nn,Chamblin:1999hg}, given by
$$\epsilon_{T}=\left(\frac{\partial \mathcal{Q}_{e}}{\partial \Phi_{e}}\right)_{T}=\frac{2 \Phi_e \Omega_2}{(a_1+a_2)^2},$$
where now the sub-index stands for at constant Hawking temperature $T$. Like in the previous situation, we note that $\epsilon_{T}$ is positive when the constant $\zeta$ and $\mu$ belong to the region ${\cal{R}}_{2}$, as shown in Figure \ref{fig:et}. Nevertheless, in this situation $\epsilon_{T} =0$ is not possible, because this implies that $a_1+a_2=0$. Curiously enough, from the intersection between the regions ${\cal{R}}_1$ and ${\cal{R}}_{2}$, the region ${\cal{R}}$ from Figure \ref{fig:termo} is naturally recovered, where this charged configuration is locally stable under thermal and electrical fluctuations.

For the sake of completeness, the extensive thermodynamic parameters (\ref{eq:mass})-(\ref{eq:charge}), and the intensive ones (\ref{eq:T}) and (\ref{eq:Phi}), fulfill the first law of black hole thermodynamics  
\begin{eqnarray}\label{eq:first-law}
\delta {\cal{M}}= T \delta {\cal{S}}+\Phi_e  \delta {\cal{Q}}_{e},
\end{eqnarray}
as well as a Smarr relation in four dimensions \cite{Smarr:1972kt,Dehghani:2013mba}
\begin{eqnarray}\label{eq:Smarr-4d}
{\cal{M}}=\frac{2}{3} \left(T {\cal{S}}+\Phi_e {\cal{Q}}_{e}\right).
\end{eqnarray}

\section{Conclusions and discussions} \label{Section-conclusions}

Due to recent research on Lovelock gravity sourced with a non minimally coupled together with self-interacting
scalar field, and its thermodynamic parameters, and the recent interest to explore an active contribution of higher gravity theories in four dimensions, in this work we have explored the four-dimensional-scalar-Einstein-Gauss-Bonnet (4DS-EGB) model (\ref{eq:EH-GB-reg}), described in \cite{Hennigar:2020lsl, Fernandes:2020nbq, Hennigar:2020fkv,Lu:2020iav,Kobayashi:2020wqy} considering planar manifolds, and adding a new matter source, consisting in a self-interacting and conformally coupled scalar field $\psi$ together with nonlinear electrodynamics characterized via a structural function ${\cal{H}}(P)$.

These configurations have only one integration constant, related to the location of the horizon $r_h$, parameterizing the metric function $f(r)$ (\ref{eq:f}) through the constant $\mu$, allowing us to switch on/off the scalar field as well as the coupling constants present in the structural function. In particular, for $\mu=1$,  ${\cal H}(P)\sim P$, while for a suitable election for the constants $\tilde{\alpha}$ and $\zeta$, we obtain the uncharged case obtained previously in \cite{BravoGaete:2013djh, Correa:2013bza}. On the other hand, for $\mu=0$ the scalar field is not present, while that ${\cal H}(P)\sim \sqrt{-2P}$. 

Additionally, with the inclusion of these matter sources in the 4DS-EGB model emerges the apparition of new interesting and non zero thermodynamic parameters,  satisfying the four-dimensional First Law (\ref{eq:first-law}) and a Smarr relation (\ref{eq:Smarr-4d}), where in order to obtain a non zero mass ${\cal{M}}$, it must exist a contribution of the scalar field and the structural function. The above shows us the importance of ${\cal{H}}(P)$, which plays a very important role in the characterization of these four-dimensional hairy charged solutions. Together with the above, and as was shown in Figure \ref{fig:termo},  it is possible to obtain positive expressions for the extensive and the intensive parameters, given a suitable election of the constants $\mu$ and $\zeta$.

It is interesting to note that this solution enjoys local stability under thermal fluctuations, thanks to the non-negativity of the specific heat $C_{\Phi_{e}}$, and under electrical fluctuation, via the positivity of the electric permittivity $\epsilon_{T}$, represented through the Figures \ref{fig:ce} and \ref{fig:et} respectively. Curiously enough, the intersection between the regions present in Figs. \ref{fig:ce}-\ref{fig:et} correspond to the sector of Fig. \ref{fig:termo},  where this charged configuration is simultaneously locally stable under thermal and electrical fluctuations. This feature has been obtained previously for Critical Gravity black holes \cite{Alvarez:2022upr}, not so for the non-linear charged configurations \cite{Bravo-Gaete:2021hza} dressed with a scalar field non-minimally coupled, where this solution enjoys local stability under thermal fluctuations but not under electrical ones.

Given that we are working on a planar base manifold, some natural open problems can arise. One of them is the possibility to explore new charged black hole solutions with some special asymptotically behavior, for example, the Lifshitz case \cite{Kachru:2008yh}, as well as the hyperscaling violation situation \cite{Charmousis:2010zz}. Together with the above, by the introduction of an improper coordinate transformation, called as Lorentz boost on a static metric (\ref{eq:metric})
\begin{equation} t\to
\frac{1}{\sqrt{1-\omega^2}}(t+\omega x_1),\,\, x_1\to
\frac{1}{\sqrt{1-\omega^2}}(x_1+ \omega t),\label{boost}
\end{equation}
which is well defined for $\omega^2<1$, we can obtain from the solution  (\ref{eq:H}),(\ref{eq:metric})-(\ref{eq:psi}), and (\ref{eq:Prt})-(\ref{sol:coupling_a}) spinning and charged configurations. Additionally, following \cite{BravoGaete:2013djh, Correa:2013bza}, it would be interesting to construct solutions for different values of the non-minimal coupling parameter $\xi$.

Finally, from a holographic motivation, this nonzero entropy ${\cal{S}}$ obtained in the present work, will allow us to explore the connection between planar black holes and the effects on shear viscosity, following the steps performed in \cite{Kovtun:2003wp,Kovtun:2004de,Son:2002sd}, where the KSS bound for the $\eta/s$ ratio can be affected due to the contribution of the coupling constant $\tilde{\alpha}$, the introduction of the conformal scalar field $\psi$, as well as the nonlinear electrodynamics through ${\cal{H}}(P)$.

\begin{acknowledgments}

 \textcolor{black}{The authors would like to thank to the anonymous referee for carefully reading our manuscript and giving valuable suggestions that led to an improved version of this work. M.B. and L.G are supported by  PROYECTO
INTERNO UCM-IN-22204, LíNEA REGULAR. J.O. also thanks the support of Proyecto de Cooperación Internacional 2019/13231-7
FAPESP/ANID, and FONDECYT REGULAR grants number 1221504 and 1210635.}

\end{acknowledgments}
\section{Appendix:} \label{Section-appendix}

\subsection{Energy-momentum tensor $ T_{\mu \nu}^{\phi}, T_{\mu \nu}^{\psi}$ and $T_{\mu\nu}^{NLE}$ from 
the Eqs.  (\ref{eq:Emunu})- (\ref{eq:EP}) }

In this subsection, we report the energy-momentum tensor $ T_{\mu \nu}^{\phi}, T_{\mu \nu}^{\psi}$ and $T_{\mu\nu}^{NLE}$ from the equations of motion present in eqs.  (\ref{eq:Emunu})-(\ref{eq:EP}) , which read
\begin{eqnarray*}
T_{\mu \nu}^{\phi}&=&4 \Big\{\frac{1}{2}\phi_{\mu}\phi_{\nu} R-2 \phi_{\lambda} \phi_{(\mu}R^{\lambda}_{\nu)}-\phi^{\lambda}\phi^{\rho} R_{\mu \lambda \nu \rho}-\phi_{\mu}^{\lambda}\phi_{\nu \lambda}\nonumber\\
&+&\phi_{\mu \nu} \Box \phi+\frac{1}{2}G_{\mu \nu}X-g_{\mu \nu}\Big[-\frac{1}{2} \phi^{\lambda \rho} \phi_{\lambda \rho}+\frac{1}{2} (\Box \phi)^2\nonumber\\
&-&\phi_{\lambda \rho} R^{\lambda \rho}\Big]\Big\}-\Big\{-4\phi R_{\mu}^{\lambda}R_{\nu \lambda}\nonumber\\
&+&2\phi g_{\mu \nu} R_{\sigma \rho } R^{\sigma \rho }+2 \phi R_{\mu \nu}R-\frac{1}{2} \phi g_{\mu \nu}R^2\nonumber\\
&-&4\phi R^{\sigma \rho} R_{\mu \sigma \nu \rho}+2\phi R_{\mu}^{\sigma \rho \tau} R_{\nu \sigma \rho \tau}-\frac{1}{2}\phi g_{\mu \nu}R_{\sigma \rho \tau \zeta} R^{\sigma \rho \tau \zeta}\nonumber\\
&-&2 R \phi_{\mu \nu}-4 R_{\mu \nu} \Box\phi+2 g_{\mu \nu} R \Box \phi +8 R_{\lambda (\mu}\phi^{\lambda}_{\nu)}\nonumber\\
&-&4 g_{\mu \nu} R_{\sigma \rho}\phi^{\sigma \rho}+4 R_{\mu (\lambda |\nu |\sigma)}\phi^{\lambda \sigma}\Big\}-4\Big[-\Box \phi \phi_{\mu} \phi_{\nu}\nonumber\\
&+& \phi_{(\mu} X_{\nu)}-\frac{1}{2} g_{\mu \nu} \phi^{\lambda} X_{\lambda}\Big]-4 X \phi_{\mu} \phi_{\nu}+ g_{\mu \nu}X^2,
\\
T_{\mu \nu}^{\psi}&=&\nabla_{\mu}\psi\nabla_{\nu}\psi - g_{\mu\nu}\Bigl[\frac{1}{2}\nabla_{\sigma}\psi\nabla^{\sigma}\psi +U(\psi)\Bigr]\nonumber \\
&+& { \xi(g_{\mu\nu}\Box -\nabla_{\mu} \nabla_{\nu}+G_{\mu\nu} )\psi^2}, 
\end{eqnarray*}
\begin{eqnarray*}
T_{\mu\nu}^{NLE} &=& \left(\dfrac{d{\cal{H}}}{dP}\right) P_{\mu\alpha}P_\nu^{\alpha} - g_{\mu\nu}\left[2P \left(\dfrac{d{\cal{H}}}{dP}\right)- {\cal{H}}\right]. 
\end{eqnarray*}

\vspace{0.1mm}

\subsection{$P^{\alpha \beta \gamma \delta}$, ${\delta \cal{L}}/{\delta (\phi_{\mu})}$,
${\delta \cal{L}}/{\delta (\psi_{\mu})}$, ${\delta \cal{L}}/{\delta (\phi_{\mu \nu})}$ and
${\delta \cal{L}}/{\delta (\partial_{\mu} A_{\nu})}$ given in ${\cal{J}}^{\mu}$  (\ref{eq:surface}) and {$Q_{(4)}$}  (\ref{eq:noether}) }

In this subsection, we present the expression of $P^{\alpha \beta \gamma \delta}$, ${\delta \cal{L}}/{\delta (\phi_{\mu})}$,
${\delta \cal{L}}/{\delta (\psi_{\mu})}$, ${\delta \cal{L}}/{\delta (\phi_{\mu \nu})}$ and
${\delta \cal{L}}/{\delta (\partial_{\mu} A_{\nu})}$ present in the surface term  (\ref{eq:surface}) as well as the $2-$form (\ref{eq:noether}):
\begin{eqnarray*}
P^{\alpha \beta \gamma \delta}&=&\frac{\delta \mathcal{L}}{\delta
R_{\alpha \beta \gamma \delta}}\nonumber\\
&=&\frac{1}{4}(1-\xi \psi^2)\,\Big(g^{\alpha\gamma}g^{\beta\delta}-g^{\alpha\delta}g^{\beta\gamma}\Big)\nonumber\\
&+&\frac{\tilde{\alpha} \phi}{2}\, R \left(g^{\alpha \gamma } g^{\beta
\delta } -g^{\alpha \delta } g^{\beta \gamma
}\right)\nonumber\\
&-&{\tilde{\alpha} \phi}\, \left(g^{\beta \delta } R^{\alpha
\gamma }-g^{\beta \gamma } R^{\alpha \delta } -g^{\alpha \delta }
R^{\beta \gamma }+g^{\alpha \gamma } R^{\beta \delta
}\right)\nonumber\\
&+&\tilde{\alpha}\phi  R^{\alpha \beta \gamma \delta}+\frac{\tilde{\alpha}}{2} \Big(g^{\alpha \gamma} \phi^{\beta}\phi^{\delta}+g^{\beta \delta} \phi^{\alpha}\phi^{\gamma}\nonumber\\
&-&g^{\beta \gamma} \phi^{\alpha}\phi^{\delta}-g^{\alpha \delta} \phi^{\beta}\phi^{\gamma}-g^{\alpha \gamma} g^{\beta \delta}X+g^{\alpha \delta} g^{\beta \gamma}X\Big),\\
\frac{\delta \cal{L}}{\delta (\phi_{\mu})}&=&4\tilde{\alpha} \Big(G^{\mu \nu} \phi_{\nu}- \Box \phi \phi^{\mu}+X \phi^{\mu}\Big),\\
\frac{\delta \cal{L}}{\delta (\psi_{\mu})}&=&-\psi^{\mu},\qquad \frac{\delta \cal{L}}{\delta (\phi_{\mu \nu})}=-2\tilde{\alpha} X g^{\mu \nu},\\
\frac{\delta \cal{L}}{\delta (\partial_{\mu} A_{\nu})}&=&-P^{\mu \nu}.
\end{eqnarray*}


\end{document}